\documentclass[letterpaper]{article}
\usepackage{aaai}
\usepackage{times}
\usepackage{helvet}
\usepackage{courier}
\newtheorem{THEOREM}{Theorem}[section]
\newenvironment{theorem}{\begin{THEOREM} \hspace{-.85em} {\bf :} }%
                        {\end{THEOREM}}
\newtheorem{LEMMA}[THEOREM]{Lemma}
\newenvironment{lemma}{\begin{LEMMA} \hspace{-.85em} {\bf :} }%
                      {\end{LEMMA}}
\newtheorem{COROLLARY}[THEOREM]{Corollary}
\newenvironment{corollary}{\begin{COROLLARY} \hspace{-.85em} {\bf :} }%
                          {\end{COROLLARY}}
\newtheorem{PROPOSITION}[THEOREM]{Proposition}
\newenvironment{proposition}{\begin{PROPOSITION} \hspace{-.85em} {\bf :} }%
                            {\end{PROPOSITION}}
\newtheorem{DEFINITION}[THEOREM]{Definition}
\newenvironment{definition}{\begin{DEFINITION} \hspace{-.85em} {\bf :} \rm}%
                            {\end{DEFINITION}}
\newtheorem{CLAIM}[THEOREM]{Claim}
\newenvironment{claim}{\begin{CLAIM} \hspace{-.85em} {\bf :} \rm}%
                            {\end{CLAIM}}
\newtheorem{EXAMPLE}[THEOREM]{Example}
\newenvironment{example}{\begin{EXAMPLE} \hspace{-.85em} {\bf :} \rm}%
                            {\end{EXAMPLE}}
\newtheorem{REMARK}[THEOREM]{Remark}
\newenvironment{remark}{\begin{REMARK} \hspace{-.85em} {\bf :} \rm}%
                            {\end{REMARK}}

\newcommand{\thm}{\begin{theorem}}
\newcommand{\lem}{\begin{lemma}}
\newcommand{\pro}{\begin{proposition}}
\newcommand{\dfn}{\begin{definition}}
\newcommand{\rem}{\begin{remark}}
\newcommand{\xam}{\begin{example}}
\newcommand{\cor}{\begin{corollary}}
\newcommand{\prf}{\noindent{\bf Proof:} }
\newcommand{\ethm}{\end{theorem}}
\newcommand{\elem}{\end{lemma}}
\newcommand{\epro}{\end{proposition}}
\newcommand{\edfn}{\bbox\end{definition}}
\newcommand{\erem}{\bbox\end{remark}}
\newcommand{\exam}{\bbox\end{example}}
\newcommand{\ecor}{\end{corollary}}
\newcommand{\eprf}{\bbox\vspace{0.1in}}
\newcommand{\beqn}{\begin{equation}}
\newcommand{\eeqn}{\end{equation}}

\newcommand{\bbox}{\vrule height7pt width4pt depth1pt}

\newcommand{\clm}{\begin{claim}}
\newcommand{\eclm}{\end{claim}}

\newcommand{\rarrow}{\rightarrow}

\newcommand{\sat}{\models}

\newcommand{\rimp}{\Rightarrow}

\newcommand{\dimp}{\Leftrightarrow}

\newcommand{\union}{\cup}
\newcommand{\inter}{\cap}

\renewcommand{\phi}{\varphi}

\newcommand{\bc}[2]{\left( \begin{array}{c} #1 \\ #2  \end{array} \right)}

\newcommand{\bd}{\bf}

\newcommand{\A}{{\cal A}}

\newcommand{\M}{{\cal M}}

\renewcommand{\P}{{\cal P}}

\newcommand{\T}{{\cal T}}

\newcommand{\V}{{\cal V}}

\newcommand{\<}{\langle}
\renewcommand{\>}{\rangle}

\newcommand{\ol}{\setlength{\itemsep}{0pt}\begin{enumerate}}
\newcommand{\eol}{\end{enumerate}\setlength{\itemsep}{-\parsep}}
\newcommand{\ul}{\setlength{\itemsep}{0pt}\begin{itemize}}
\newcommand{\dl}{\setlength{\itemsep}{0pt}\begin{description}}
\newcommand{\edl}{\end{description}\setlength{\itemsep}{-\parsep}}
\newcommand{\eul}{\end{itemize}\setlength{\itemsep}{-\parsep}}

\newcommand{\true}{\mbox{{\it true}}}
\newcommand{\false}{\mbox{{\it false}}}

\newcommand{\commentout}[1]{}

\newcommand{\bi}{\begin{itemize}}
\newcommand{\ei}{\end{itemize}}
\newcommand{\be}{\begin{enumerate}}
\newcommand{\ee}{\end{enumerate}}

\newcommand{\intension}[1]{[\![ #1 ]\!]}

\newcommand{\satp}{\models^{\, \mathit{sp}}}
\renewcommand{\L}{{\cal L}}

\renewcommand{\P}{{\cal P}}
\newcommand{\vdashL}{\vdash_{\Lambda}}

\newcommand{\Deltafo}{\Delta_{\mathit{fo}}}
\newcommand{\Deltar}{\Delta_{\rarrow}}
\newcommand{\tendsto}{\rightarrow}
\newcommand{\LCondfo}{\L_C}
\newcommand{\Lfo}{\L^{\mathit{fo}}}
\newcommand{\LCondfoo}{\L_C^-}
\newcommand{\LCondfoq}{\L_{C,q}}
\newcommand{\AXC}{\mbox{AX}_C}
\newcommand{\AXCL}{\mbox{AX}_C^\Lambda}
\newcommand{\Cond}{\mbox{\boldmath$\rightarrow$\unboldmath}}
\newcommand{\PBox}{N}

\newcommand{\DIST}{\mathit{Dist}}
\newcommand{\dist}{\mathit{dist}}
\newcommand{\Pfo}{\mathbf{P}_\Lambda^+}
\newcommand{\Pfoq}{\mathbf{P}_\Lambda^{+,q}}
\newcommand{\Pfop}{\mathbf{P}^{+}_{\Lambda}}
\newcommand{\Pfopq}{\mathbf{P}_\Lambda^{+,q}}
\newcommand{\hra}{\hookrightarrow}
\renewcommand{\bd}{\mathbf{d}}
\renewcommand{\bc}{\mathbf{c}}
\newcommand{\bD}{\mathbf{D}}
\newcommand{\fullv}[1]{#1}
\newcommand{\shortv}{\commentout}

\newcommand{\PS}{\mathcal{PS}}
\newcommand{\sysPL}{\mathbf{P}_{\Lambda}}
\newcommand{\sysPLD}{\mathbf{P}_{\Lambda \union \Delta^*_0}}

\begin{document}

\title{From Qualitative to Quantitative Proofs of Security Properties\\
Using First-Order Conditional Logic}

\author{Joseph Y. Halper%
\thanks{Supported in part by NSF under
under grants ITR-0325453 and IIS-0534064, and by AFOSR under grant 
FA9550-05-1-0055.}\\
Cornell University\\
Dept. of Computer Science\\
Ithaca, NY 14853\\
halpern@cs.cornell.edu\\
http://www.cs.cornell.edu/home/halpern}
\date{}

\maketitle

\begin{abstract}
A first-order conditional logic is considered, with semantics given by a
variant of 
\emph{$\epsilon$-semantics} \cite{Adams:75,Goldszmidt92}, where $\phi
\Cond \psi$ means that $\Pr(\psi \mid \phi)$ approaches 1
\emph{super-polynomially}---faster than any inverse polynomial.  This
type of convergence is needed for reasoning about  
security protocols.  A complete axiomatization is provided for this
semantics, and it is   
shown how a  qualitative proof of the correctness of a security protocol
can be automatically converted to a quantitative proof appropriate for
reasoning about concrete security.
\end{abstract}

\section{Introduction}

Security protocols, such as key-exchange and key-management protocols,
are short, but  
notoriously difficult to prove correct.  
\shortv{
To give just one of many examples, several security flaws have been
found in the 802.11 Wired Equivalent Privacy (WEP) protocol used to protect
link-layer communications from eavesdropping and other attacks
\cite{BGW2001}.}  
\fullv{Flaws have been found in numerous protocols, ranging from the 
the 802.11 Wired Equivalent Privacy (WEP) protocol used to protect
link-layer communications from eavesdropping and other attacks
\cite{BGW2001} to standards and proposed standards for Secure Socket
Layer~\cite{WS1996,MSS1998} to Kerberos \cite{BP1998}.}
Not surprisingly, a great deal of effort has been devoted to proving the
correctness of such protocols.  There are two largely disjoint
approaches.  The first
essentially ignores the details of cryptography by assuming perfect
cryptography (i.e., nothing encrypted can ever be decrypted without the
encryption key) and an adversary that controls the network.  By ignoring
the cryptography, it is possible to give a more qualitative proof of
correctness, using logics designed for reasoning about security
protocols.  Indeed, this approach has enabled axiomatic proofs of
correctness and model checking of proofs (see, for example,
\cite{r:mitchell97,r:paulson94}). 
The second approach applies the tools of 
modern cryptography to proving correctness, using more quantitative
arguments.  Typically it is shown that, 
given some security parameter $k$ (where $k$ may
be, for example, the length of the key used) an adversary whose running
time is polynomial in $k$ has a negligible probability of breaking the
security, where ``negligible'' means ``less than any inverse polynomial
function of $k$'' \shortv{(see, for example, \cite{goldreich01}).}
\fullv{(see, for example, \cite{BCK1998,goldreich01}).}

There has been recent work on bridging the gap between these two approaches,
with the goal of constructing a logic that can allow reasoning
about quantitative aspects of security protocols while still being
amenable to mechanization.   This line of research started with the
work of Abadi and Rogaway \citeyear{r:abadi00}.  More recently, Datta et
al. \citeyear{DDMST05} showed that by giving a somewhat nonstandard
semantics to 
their first-order \emph{Protocol Composition Logic}
\cite{DDMR07}, it was possible to reason about many features of the
computational model.  In this logic, an ``implication'' of the form
$\phi \supset B$ is 
interpreted as, roughly speaking, the probability of $B$ given $\phi$ is
high.  For example, a statement like {\tt secret
encrypted} $\supset$ {\tt adversary does not decrypt the secret} says
``with high 
probability, if the secret is encrypted, the adversary does not decrypt
it''.  While the need for such statements should be clear, 
the probabilistic interpretation used is somewhat unnatural, and no
axiomatization is provided by Datta et al. \citeyear{DDMST05} for the
$\supset$ operator (although some sound axioms are given that use it).

The interpretation of $\supset$ is quite reminiscent of one of the
interpretations of $\rarrow$ in conditional logic, where $\phi \rarrow \psi$
can be interpreted as ``typically, if $\phi$ then $\psi$'' \cite{KLM}.  
Indeed, one semantics given to $\rarrow$, called \emph{$\epsilon$-semantics} 
\cite{Adams:75,Goldszmidt92}, is very close in spirit to that used in
\cite{DDMST05}; this is particularly true for the formulation of
$\epsilon$-semantics given by Goldszmidt, Morris, and Pearl
\citeyear{GMPfull}.  In this formulation, a formula  $\phi \rarrow \psi$ is
evaluated with respect to a sequence
$(\Pr_1, \Pr_2, \ldots)$ of probability measures (\emph{probability
sequence}, for short): it is true if, roughly
speaking, \shortv{$\lim_{n\rarrow \infty} \Pr_n(\psi \mid \phi) = 1$.}
\fullv{$\lim_{n\rarrow \infty} \Pr_n(\psi \mid \phi) = 1$ 
(where $\Pr_k(\psi \mid \phi)$ is taken to be 1 if $\Pr_k(\phi) =
1$).}   
This formulation is not
quite strong enough for some security-related purposes, where the
standard is 
\emph{super-polynomial} convergence, that is, convergence
faster than any inverse polynomial.  To capture such convergence, we can
take
$\phi \rarrow \psi$ to 
be true with respect to this probability sequence if, for all
polynomials $p$, there 
exists $n^*$ such that, for all $n \ge n^*$, $\Pr_n(\psi \mid \phi) \ge 1 -
1/p(n)$.  (Note that this implies that $\lim_{n \rarrow \infty}
\Pr_n(\psi \mid \phi) = 1$.)  
\fullv{In a companion paper,} 
\shortv{
In a companion paper \cite{DHMPR08},} 
it is shown that reinterpreting $\rarrow$ in this way gives an
elegant, powerful variant of the logic considered in \cite{DDMST05},
which can be used to reason about security protocols of interest.

While it is already a pleasant surprise that conditional logic provides
such a clean approach to reasoning about security, using conditional
logic has two further significant advantages, which are the subject
of this paper.  The first is that, as I show here,
the well-known complete axiomatization of conditional logic with respect
to $\epsilon$-semantics continues to be sound and complete with respect
to the super-polynomial semantics for $\rarrow$; thus, the
axioms form a basis for automated proofs.  The second is that the use of
conditional logic allows for a clean transition from
qualitative to quantitative arguments.  To explain these points, I need
to briefly recall some well-known results from  the literature.  

As is well known, the \emph{KLM properties}
\cite{KLM} (see Section~\ref{sec:review}) provide 
a sound and complete axiomatization for reasoning about
$\rarrow$ formulas with respect to $\epsilon$-semantics \cite{Geffner92}.
More precisely, if $\Delta$ is a collection of formulas of the form
$\phi' \rarrow \psi'$, then $\Delta$ \mbox{\emph{($\epsilon$-)entails}} $\phi
\rarrow \psi$ 
(that is, for every probability sequence $\P$, if every formula in
$\Delta$ is true in $\P$ according to $\epsilon$ semantics, then so is
$\phi \rarrow \psi$), then $\phi \rarrow \psi$ is provable from $\Delta$
using the KLM properties.   
This result applies only when $\Delta$ is a collection of $\rarrow$ formulas.
$\Delta$ cannot include negations 
or disjunctions of $\rarrow$ formulas.
\emph{Conditional logic} extends the KLM framework by allowing Boolean
combinations of $\rarrow$ statements.  A sound and complete axiomatization of
propositional
conditional logic with semantics given by what are called preferential
structures was given by Burgess \citeyear{Burgess81}; Friedman and
Halpern \citeyear{FrH5full} proved it was also sound and complete for
$\epsilon$-semantics.  

Propositional conditional logic does not suffice for reasoning about
security. 
The logic of \cite{DDMST05} is first-order; quantification is needed to
capture important properties of security protocols.  A sound and
complete axiomatization for the language of first-order conditional
logic, denoted $\LCondfo$, with respect to
$\epsilon$-semantics is given by Friedman, Halpern, and Koller
\citeyear{FHK}.  
The first major result of this paper shows
a conditional logic formula $\phi$ is satisfiable in some model
$M$ with respect to $\epsilon$-semantics iff it is satisfiable in some
model $M'$ with respect to the super-polynomial semantics.  It follows
that all the completeness results for $\epsilon$-semantics apply without
change to the super-polynomial semantics.

I then consider the language $\LCondfo^0$
which essentially consists of universal $\rarrow$ formulas, 
that is, formulas of 
the form $\forall x_1 \ldots
\forall x_n (\phi \Cond \psi)$, where $\phi$ and $\psi$ are first-order
formulas. 
As in the KLM framework, there are no nested
$\rarrow$ formulas or negated $\rarrow$ formulas. 
The second major result of this paper is to provide a sound and complete
axiomatization that extends the KLM properties
for reasoning abut when a collection  
of formulas in $\LCondfo^0$ entails a formula in $\LCondfo^0$.

It might seem strange to be interested in an
axiomatization for $\LCondfo^0$ when there is already a sound and
complete axiomatization for the full language $\LCondfo$.
However, $\LCondfo^0$ has some significant advantages.
In reasoning about concrete security, asymptotic complexity results do
not suffice; more detailed information about security
guarantees is needed.  For example, we may want to prove that an SSL
server that supports 1,000,000 sessions using 1024 bit keys has a
probability of 
0.999999 of providing the desired service without being compromised.  I
show how to convert a 
qualitative proof of security in  the language $\LCondfo^0$,
which provides only asymptotic guarantees, to a quantitative proof.
Moreover, the conversion shows exactly how strong the assumptions have to
be in order to get the desired 0.999999 level of security.
Such a conversion is not posisble with $\LCondfo$.

This conversion justifies reasoning at the qualitative level.  A
qualitative proof can be constructed without worrying about the details
of the numbers, and then automatically converted to a quantitative proof
for the desired level of security. 

\fullv{
In the next section, I review the syntax and semantics of conditional
logic, with an emphasis on $\epsilon$ semantics, and 
show how it can be modified to deal with the super-polynomial 
convergence that is more appropriate for reasoning about security.
In Section~\ref{sec:axiomatization}, I provide axioms and inference
rules for both 
qualitative and quantitative reasoning.  
\shortv{All missing proofs can be found in the full paper 
(available at
http://www.cs.cornell.edu/home/halpern/papers/quantsecurity.pdf) 
}
}

\section{First-Order Conditional Logic}\label{sec:review}

I review the syntax and semantics of first-order conditional logic here.
Although I focus on first-order conditional logic here, it is
straightforward to specialize all the definitions and results to the
propositional case, so I do not discuss the propositional case further.

The syntax of first-order conditional logic is straightforward.  
Fix a finite first-order \emph{vocabulary} $\T$ consisting, as usual,
of function 
symbols, predicate symbols, and constants.  Starting with atomic
formulas of first-order logic over the vocabulary $\T$, more
complicated formulas are formed by closing off under the standard
truth-functional 
connectives 
(i.e., $\land$ ,$\lor$, $\neg$, and $\rimp$), first-order quantification,
and the binary modal operator $\Cond$.  Thus, a typical formula is 
$\forall x (P(x) \Cond \exists y (Q(x,y) \Cond R(y)))$.
Let $\LCondfo(\T)$ be the resulting language.  
Let $\Lfo(\T)$ be the pure first-order fragment of $\LCondfo(\T)$,
consisting of $\rarrow$-free formluas.
Let $\LCondfo^0(\T)$ consist
of all formulas in $\LCondfo(\T)$ of the form $
\forall x_1 \ldots \forall x_n ( \phi \Cond \psi)$, where
$\phi$ and $\psi$ are in $\Lfo$.
(I henceforth omit the $\T$ unless it is necessary for clarity.)
Note that $\LCondfo^0$ does not include negations of $\rarrow$
formulas or conjunctions of $\rarrow$ formulas.
While not having conjunctions does not really impair the expressive
power of $\LCondfo^0$ (since we will be interested in sets of
$\LCondfo^0$ formulas, where a set can be
identified with the conjunction of the formulas in the set), 
the lack of negation does.

I give two semantics to formulas in $\LCondfo(\T)$.  In both semantics,
the truth of formulas is defined with respect to \emph{PS structures}.
A 
PS structure is a tuple $M = (D, W,\pi, \P)$, 
where $D$ is a domain, $W$ 
is a set of worlds, $\pi$ is an \emph{interpretation}, which associates
with each predicate symbol (resp., function symbol, constant symbol) in
$\T$ and world $w \in W$ a predicate (resp., function, domain element)
of the right arity, and $\P = \<\Pr_1, \Pr_2, \ldots\>$ is a probability
sequence.  As
usual, a \emph{valuation} $V$ 
associates with each variable $x$ an element $V(x) \in D$.  

Given a valuation $V$ and structure $M$, the semantics of $\land$,
$\neg$, $\rimp$, and $\forall$ is completely standard.  In particular,
the truth of a first-order formula in $\Lfo$ in a world $w$, written 
$(M,V,w) \models \phi$, is determined as usual.  For $\phi \in
\Lfo$, let
$\intension{\phi}_{M,V} = \{w: (M,V,w) 
\models \phi\}$.  
If $\phi$ is a closed formula, so that its truth does
not depend on the valuation, I occasionally write $\intension{\phi}_M$
rather than $\intension{\phi}_{M,V}$.
I write $(M,V) \sat \phi$ if $(M,V,w) \sat \phi$ for all worlds $w$.
The truth of an $\rarrow$ formula does not depend on
the world, but only on the structure $M$.  
$$(M,V) \models \phi \Cond \psi \mbox{ if } \lim_{n \rightarrow \infty}
{\Pr}_n (\intension{\psi}_{M,V}  \mid \intension{\phi}_{M,V}) = 1,$$
where ${\Pr}_n (\intension{\psi}_{M,V}  \mid \intension{\phi}_{M,V})$ is
taken to be 1 if ${\Pr}_n(\intension{\phi}_{M,V}) = 0$.

I also consider an alternative semantics that gives
super-polynomial convergence.  A polynomial is
\emph{positive} if all its coefficients are nonnegative and at least one
is nonzero.  
$$\begin{array}{ll}
(M,V) \satp \phi \Cond \psi \mbox{ if for all positive polynomials $p$,}\\
\mbox{there exists some $n^* \ge 0$
such that, for all $n \ge n^*$},\\
{\Pr}_n (\intension{\psi}_{M,V}  \mid \intension{\phi}_{M,V}) \ge 1 - (1/p(n)).
\end{array}$$

As usual, I write $M \sat \phi$ if $(M,V) \sat \phi$ for all valuations
$V$, 
 and $\M \sat \phi$ if $M \sat \phi$ for all PS structures in a set $\M$,
and similarly with $\sat$ replaced by $\satp$.  

\section{Axioms for qualitative and quantitative
reasoning}\label{sec:axiomatization}   
In this section, I start by showing that qualitative reasoning for
both $\sat$ and $\satp$ is characterized by the same axiom system.
I then provide a complete axiomatization for 
$\LCondfo^0$.  Finally, I consider quantitative conditional logic.
In the axioms, it is convenient to use $\PBox \phi$ as an abbreviation for
$\neg \phi \Cond \false$.  Note that if $\phi$ is a closed formula, then
$M \sat \PBox \phi$ iff, for some $n^*$, 
${\Pr}_n(\intension{\phi}_M) = 0$ for all $n \ge n^*$, and similarly with
$\sat$ replaced by $\satp$.  Thus, $\PBox \phi$ can be read as saying
``$\phi$ is almost surely true''.

\subsection{Qualitative Reasoning}
As was mentioned in the introduction, Friedman, Halpern, and Koller
\citeyear{FHK} provide a complete axiomatization $\AXC$ for $\LCondfo$
with respect to $\sat$.  
For the security applications, a generalization
of their result is needed, where it is possible to
restrict to models where all 
worlds satisfy a particular first-order theory $\Lambda$.
Let $\vdashL$ denote provability in first-order logic given the axioms
in the theory $\Lambda$.
Let $\AXCL$ consist of the following axioms and rules:

\begin{description}
\item[$\Lambda$-AX.] $\phi$, if $\phi \in \Lfo$ and
$\vdashL \phi$.
\item[C0.] All
substitution
instances of propositional tautologies.
\item[C1.] $\phi \Cond \phi$.
\item[C2.]
$((\phi \Cond \psi_1)\land(\phi\Cond\psi_2)) \rimp
    (\phi\Cond(\psi_1\land\psi_2))$.
\item[C3.]
$((\phi_1\Cond\psi)\land(\phi_2\Cond\psi)) \rimp
    ((\phi_1\lor\phi_2) \Cond\psi)$.
\item[C4.]
$((\phi_1\Cond\phi_2)\land(\phi_1\Cond\psi)) \rimp
    ((\phi_1\land\phi_2) \Cond \psi)$.
\item[C5.]
$[(\phi \Cond \psi) \rimp \PBox(\phi \Cond \psi)] \land
            [\neg(\phi \Cond \psi) \rimp \PBox \neg (\phi \Cond \psi)]$.
\item[C6.] $\neg (\true \Cond \false)$.
\item[F1.] $\forall x \phi \rimp \phi[x/t]$,
where $t$ is {\em substitutable\/} for $x$ in the sense discussed below
and $\phi[x/t]$ is the result of substituting $t$ for all free
occurrences of $x$ in $\phi$ (see \cite{Enderton} for a formal
definition).
\item[F2.] $\forall x(\phi \rimp \psi) \rimp (\forall x \phi
\rimp \forall x \psi)$.
\item[F3.] $\phi \rimp \forall x \phi$ if $x$ does not occur free in
$\phi$.
\item[F4.] $x = y \rimp (\phi_1 \rimp \phi_2)$, where $\phi_1$ is
quantifier-free
and $\phi_2$ is obtained
{f}rom $\phi_1$ by replacing zero or more occurrences of $x$ in $\phi_1$
by $y$.
\item[F5.] $x \ne y \rimp \PBox(x \ne y)$.
\item[MP.] From $\phi$ and $\phi \rimp \psi$ infer $\psi$.
\item[Gen.] From $\phi$ infer $\forall x \phi$.
\item[R1.] From $\phi_1 \dimp \phi_2$
infer $\phi_1\Cond\psi \dimp \phi_2\Cond\psi$.
\item[R2.] From $\psi_1 \rimp \psi_2$ infer $\phi\Cond\psi_1 \rimp
    \phi\Cond\psi_2$.
\end{description}

The axiom system $\AXC$ of \cite{FHK} does not have $\Lambda$-AX (this is
needed to incorporate the theory $\Lambda$) and includes an axiom $x=x$
that follows from $\Lambda$-AX; otherwise, the axiom systems are identical.
\fullv{
As observed in \cite{FHK}, the ``positive'' version of F5,
$x = y \rimp \PBox(x=y)$, is also sound.  It is not included in the
axiomatization because it is provable from the other axioms.}

It remains to explain the notion of ``substitutable'' in F1.
Clearly a term $t$ with free variables
that might be captured by some quantifiers in $\phi$ cannot be
substituted for $x$; for example,
while $\forall x \exists y (x \ne y)$ is true as long as the domain
has at least two elements, the result of substituting $y$ for $x$ is
$\exists y (y \ne y)$, which is surely false.  In the case of
first-order logic, it suffices to define ``substitutable'' so as
to make sure this does not happen (see
\cite{Enderton} for details).  However, in modal logics such as this one,
more care must be taken
In general, terms cannot be substituted for universally
quantified variables in a modal context, since terms are not
in general \emph{rigid}; that is, they can have different
interpretations in different worlds.  To understand the impact of this,
consider the formula $\forall x (\neg \PBox P(x))
\rimp \neg \PBox P(c)$ (where $P$ is a unary predicate and $c$ is a
constant).  This formula is not valid in PS structures.  For example,
consider a PS structure with two worlds $w_1$ and $w_2$, and a domain
with two elements $d_1$ and $d_2$.  Suppose that in world $w_1$,
$P(d_1)$ holds, $P(d_2)$ does not, and $c$ is interpreted as $d_1$,
while in world $w_2$, $P(d_2)$ holds, $P(d_1)$ does not, and $c$ is
interpreted as $d_2$.  Then it is easy to see that  $\PBox P(c)$ holds
in both worlds, but $\PBox P(x)$ holds in only one world, no matter how $x$
is interpreted.  If 
$Pr_n(w_1) = Pr_n(w_2) = 1/2$ for all $n$, then $M \sat \PBox P(c)$,
while $M \sat \forall x (\neg \PBox P(x))$.  Thus, 
if $\phi$ is a formula that has occurrences of $\Cond$, then
the only terms that are considered substitutable for $x$ in $\phi$
are other variables.

\fullv{
It is interesting to contrast R1, R2, and C4.  While R2 allows a formula
$\psi_1$ on the right-hand side of $\Cond$ to be replaced by a weaker
formula $\psi_2$ (that is, a formula such that that $\psi_1 \rimp \psi_2$ is
provable), R1 just allows a formula $\phi_1$ on the left-hand
side of $\Cond$ to be replaced by an equivalent formula $\phi_2$, rather
than a stronger formula.  C4 allows the replacement of a formula
$\phi_1$ on 
the left-hand side by a stronger formula, $\phi_1 \land \phi_2$, but
only if $\phi_1 \Cond \phi_2$ holds.
Intuitively, this says that if $\phi_2$ and $\psi$ each almost always
hold given $\phi_1$, then $\psi$ almost always holds given both $\phi_1$
and $\phi_2$.  Monotonicity does not hold in general.  
That is, if if $\phi_1 \rimp \phi_2$ is provable and $\phi_2 \Cond \psi$
holds, then $\phi_1 \Cond \psi$ does not necessarily hold.
For a simple
counterexample, it is not the case that if $\true \Cond \psi$ holds then $\neg
\psi \Cond \psi$ holds.  If $\psi(x)$ 
states that $x$ cannot break the encryption, it seems reasonable to
expect that, almost
always, $x$ cannot break the encryption ($\true \Cond \psi(x)$), but it
surely is not the case that $x$ cannot break the encryption given
that $x$ can break it. 

The remaining axioms and rules are easy to explain.  In particular, 
C2 says that if both $\psi_1$ and $\psi_2$ almost always hold given
$\phi$, then so does $\psi_1 \land\psi_2$, while C3 allows reasoning by
cases: if $\psi$ almost always holds given each of $\phi_1$ and
$\phi_2$, then it almost always holds given their disjunction.  
}

I want to show that $\AXCL$ is also sound and complete for the $\satp$
semantics.  
The key step in doing that is 
to show that a formula is satisfiable with respect to the
$\sat$ semantics iff it is satisfiable with respect to the $\satp$
semantics.  

\begin{theorem}\label{thm:equisatisfiable} If $M = (D, W, \pi,\P)$ is a PS
structure and $D$ is countable, 
then there exists a probability sequence $\P'$ such that,
for all valuations $V$,
$(M,V) \sat \phi$ iff $(M',V) \satp \phi$, where
$M' = (D,W,\pi,\P')$.
\end{theorem}

\prf
Suppose that $M = (D,W,\pi,\P)$, where $D=\{d_1, d_2, \ldots\}$
($D$ may be finite), and $\P = (\Pr_1, \Pr_2, \ldots)$. 
Let $L = (\phi_1 \Cond \psi_1, \phi_2 \Cond \psi_2, \ldots)$ be a list
of all formulas of the form $\phi' \Cond \psi'$ in 
$\LCondfo$ with the property that if $(M,V') \sat \neg(\phi' \Cond
\psi')$ for some valuation $V'$, 
then $\phi' \Cond \psi'$ appears infinitely often in $L$.
Suppose that the set of variables is
$\{x_1, x_2, \ldots\}$.  (I am implicitly assuming that the set of
variables is countable, as is standard.)
Let $\V_n$ be the set of valuations $V$ such that $V(x_i) \in \{d_1,
\ldots, d_n\}$ for $i = 1, \ldots, n$ 
and $V(x_m) = d_1$ for all $m > n$.   Given a valuation $V'$ and a formula
$\phi \in \LCondfo$, there exists $n$ such that, for all free variables
$x$ in $\phi$, 
$x \in \{x_1, \ldots, x_n\}$ and $V'(x) \in \{d_1, \ldots, d_n\}$.
Thus,
$(M,V') \sat \phi$ for some valuation 
$V'$ iff $(M,V') \sat \phi$ for some valuation $V' \in \V_n$.
Suppose that the elements of $\V_n$ are $V^n_1, \ldots, V^n_{|\V_n|}$.

Since $\V_n$ is finite, there is
a subsequence $\P' = (\Pr_{11}', \ldots
\Pr_{1|\V_1|},  \Pr_{21}', \ldots, \Pr_{2|\V_2|}, \ldots)$ of $\P$
with the following properties, for $1 \le m \le |\V_n|$:  
\begin{equation}\label{eqprop1}
\begin{array}{l}
\mbox{for all $j \le n$ and $V' \in \V_n$, 
if $(M,V') \sat \phi_j \Cond \psi_j$,}\\
\mbox{then $\Pr'_{nm}(\intension{\psi_j}_{M,V'} \mid
\intension{\phi_j}_{M,V'}) \ge 1 - 1/n^n$;}
\end{array}
\end{equation}
\begin{equation}\label{eqprop2}
\begin{array}{l}
\mbox{if $(M,V^n_m) \sat \neg(\phi_n \Cond \psi_n)$, then}\\ 
\mbox{$\Pr'_{nm}(\intension{\psi_n}_{M,V^n_m} \mid
\intension{\phi_n}_{M,V^n_m}) < 1 - 1/k$, where $k$}\\ 
\mbox{is the smallest
integer such that, for infinitely many}\\ 
\mbox{indices $h$, $\Pr_h(\intension{\psi_n}_{M,V^n_m} \mid
\intension{\phi_n}_{M,V^n_m}) < 1 - 1/k$.}
\end{array}
\end{equation}
(There must be such a $k$, since $\lim_{h \tendsto \infty} 
\Pr_h(\intension{\psi_n}_{M,V^n_m} \mid \intension{\phi_n}_{M,V^n_m}) \ne 1$.)

Let $M' - (D, W, \pi, \P')$.  
I now prove that $(M,V) \sat \phi$ iff $(M',V) \satp \phi$ 
for all valuations $V$ and formulas $\phi \in \LCondfo$ 
by a straightforward induction on the
structure of $\phi$.  If $\phi$ is an atomic formula, this is immediate,
since $M$ and $M'$ differ only in their probability sequences.  All
cases but the one where $\phi$ has the form $\phi' \Cond \psi'$ follow
immediately from the induction hypothesis.  If $\phi$ has the form $\phi'
\Cond \psi'$, first suppose that $(M,V) \sat \phi' \Cond \psi'$.  Fix a
polynomial $p$.  There must exist some $n^*$ such that 
(a) for all free variables $x$ in $\phi'$ or $\psi'$, $x \in \{x_1, \ldots,
x_{n^*}\}$ and $V(x) \in \{d_1, \ldots d_{n^*}\}$,  
(b) $p(n) < 1/n^n$ for all $n \ge n^*$, and (c) $\phi' \Cond \psi'$
is among the first $n^*$ formulas in $L$.   It follows from (a) that for
all $n \ge n^*$, there exists
some $V' \in \V_N$ such that $V'$ and $V$ agree on all the free
variables in $\phi' \Cond \psi'$.  
It then follows from (b), (c), and (\ref{eqprop1}) that, for all $n \ge
n^*$ and $1 
\le m \le |\V_n|$, 
$\Pr'_{nm}(\intension{\psi'}_{M,V} \mid
\intension{\phi'}_{M,V}) \ge 1 - 1/p(n)$.
Thus, $(M,V) \satp \phi' \Cond \psi'$.

If $(M,V) \sat \neg(\phi' \Cond \psi')$, there must be
some minimal $k$ such that  $\Pr_h(\intension{\psi'}_{M,V} \mid
\intension{\phi'}_{M,V}) < 1 - 1/k$ for infinitely many indices $h$.
Since $\phi' \Cond \psi'$ occurs infinitely often in $L$, it easily
follows from (\ref{eqprop2}) that, for infinitely many values of $n$ and
$h$, 
$\Pr'_{nh}(\intension{\psi'}_{M,V} \mid
\intension{\phi'}_{M,V}) < 1 - 1/k$.
Let $p(n) =
k$ (so $p(n)$ is a constant function).  It follows that
$\Pr'_{nh}(\intension{\psi'}_{M,V} \mid
\intension{\phi'}_{M,V}) < 1 - 1/p(n)$ for infinitely many values of $n$
and $h$.
Thus, $(M,V) \satp \neg(\phi' \Cond \psi')$.
This completes the proof. \eprf

Let $\PS(\Lambda)$ consist of all PS structures $M$ where every world
satisfies $\Lambda$.  

\begin{theorem}\label{thm:soundandcomplete1}  $\AXCL$ is a sound and
complete axiomatization for $\PS(\Lambda)$ 
with respect to both $\sat$ and
$\satp$.  That is, the following are equivalent for all formulas in
$\LCondfo(\T)$: 
\begin{description}
\item[{\it (a)}] $\AXCL \vdash \phi$;
\item[{\it (b)}] $\PS(\Lambda) \sat \phi$;
\item[{\it (c)}] $\PS(\Lambda) \satp \phi$.
\end{description}
\end{theorem}

\prf The equivalence of parts (a) and (b) for the case that $\Lambda =
\emptyset$ is proved in Theorem
5.2 of \cite{FHK}.  The same proof shows that the result holds for
arbitrary $\Lambda$.  To show that (a) implies (c), I must show that
all the axioms are sound.
The soundness of all the axioms and rules other than C2, C3, C4, and C5
is trivial.  I consider each of these axioms in turn.

For C2, suppose that $M = (D, W,\pi, \<\Pr_1, \Pr_2, \ldots\>)$ is a
PS structure such that $M \satp \phi \Cond \psi_1$ and $M \satp \phi
\Cond \psi_2$.  Since $M \satp \phi \Cond \psi_i$, $i = 1, 2$, 
given a positive polynomial $p$, 
there exists $n_1^*, n_2^* \ge 0$ such that, for all $n \ge n_i^*$, 
${\Pr}_n(\intension{\psi_i}_{M,V} \mid \intension{\phi}_{M,V}) \ge 1 -
1/2p(n)$, for $i = 1, 2$.  For all $n \ge \max(n_1^*,n_2^*)$, 
$\Pr_n(\intension{\psi_i} \mid \intension{\phi}) \le 1/2p(n)$.  Thus, 
for $n \ge \max(n_1^*, n_2^*)$, 
$$\begin{array}{lll}
&{\Pr}_n(\intension{\psi_1 \land \psi_2}_{M,V} \mid 
\intension{\phi}_{M,V})\\
\ge & 1 - (\Pr_n(\intension{\psi_1} \mid 
\intension{\phi}) + \Pr_n(\intension{\psi_2} \mid \intension{\phi}))\\
\ge &1 - \frac{1}{2p(n)} - \frac{1}{2p(n)} \\
= &1 - \frac{1}{p(n)}.
\end{array}
$$
\shortv{The proof of soundness for the remaining axioms can be found in
the full paper.}

\fullv{
For C3, note that 
\begin{equation}\label{eq:or}
\begin{array}{ll}
&\Pr(A \mid B_1 \union B_2) \\
= &\Pr((A \inter B_1 \union A \inter B_2)\mid B_1 \union B_2) \\
= &\Pr(A \inter B_1 \mid B_1 \union B_2) + 
\Pr(A \inter B_2 \mid B_1 \union B_2) \\
&-  
\Pr(A \inter B_1 \inter B_2 \mid B_1 \union B_2)\\
= &\Pr(A \mid B_1) \times \Pr(B_1 \mid B_1 \union B_2)\\
& +  
\Pr(A \mid B_2) \times \Pr(B_2 \mid B_1 \union B_2) \\
&- \Pr(A \inter B_1 \inter B_2 \mid B_1 \union B_2).
\end{array}
\end{equation}
Now suppose that $M \satp \phi_1 \Cond \psi$ and $M \satp \phi_2
\Cond \psi$.
Given a positive polynomial $p$, as in the case of C2, there exist 
$n_1^*$ and $n_2^*$ such that,
for all $n \ge n_i^*$, 
${\Pr}_n(\intension{\psi}_{M,V} \mid \intension{\phi_1}_{M,V}) \ge 1 -
1/2p(n)$, for $i = 1, 2$.
It easily follows from (\ref{eq:or}) that if $n \ge \max(n_1^*, n_2^*)$,
then 
$$\begin{array}{ll}
&{\Pr}_n(\intension{\psi}_{M,V} \mid \intension{\phi_1 \lor
\phi_2}_{M,V})\\
 \ge &(1 - \frac{1}{2p(n)}) {\Pr}_n(\intension{\phi_1}_{M,V} \mid
\intension{\phi_1 \lor \phi_2}_{M,V}) \\
&+ (1 - \frac{1}{2p(n)}) {\Pr}_n(\intension{\phi_2}_{M,V} \mid
\intension{\phi_1 \lor \phi_2}_{M,V})\\ & - {\Pr}_n(\intension{\psi \land
\phi_1 \land \phi_2}_{M,V} \mid \intension{\phi_1 \lor \phi_2}_{M,V})\\
 \ge &(1 - \frac{1}{2p(n)}) {\Pr}_n(\intension{\phi_1}_{M,V} \mid
\intension{\phi_1 \lor \phi_2}_{M,V}) \\
&+ (1 - \frac{1}{2p(n)}) {\Pr}_n(\intension{\phi_2}_{M,V} \mid
\intension{\phi_1 \lor \phi_2}_{M,V})\\ & - {\Pr}_n(\intension{\phi_1 \land
\phi_2}_{M,V} \mid \intension{\phi_1 \lor \phi_2}_{M,V})\\ 
\ge &(1 - \frac{1}{2p(n)}) [{\Pr}_n(\intension{\phi_1}_{M,V} \mid
\intension{\phi_1 \lor \phi_2}_{M,V})\\ &+ 
{\Pr}_n(\intension{\phi_2}_{M,V} \mid
\intension{\phi_1 \lor \phi_2}_{M,V})\\ & - 
{\Pr}_n(\intension{\phi_1 \land \phi_2}_{M,V} \mid
\intension{\phi_1 \lor \phi_2}_{M,V})]\\& - 
\frac{1}{2p(n)}{\Pr}_n(\intension{\phi_1 \land \phi_2}_{M,V} \mid
\intension{\phi_1 \lor \phi_2}_{M,V}) \\ 
\ge &(1 - \frac{1}{2p(n)}) - \frac{1}{2p(n)}\\
= &1 - \frac{1}{p(n)}.
\end{array}
$$

For C4, note that 
$$
\begin{array}{ll}
&\Pr(A_1 \mid A_2 \inter B)\\
 = &\Pr(A_1 \inter A_2 \mid
B)/\Pr(A_2 \mid B) \ge \Pr(A_2 \inter A_2 \mid B),
\end{array}$$ 
so the argument follows essentially the same lines as that for C2.

Finally, C5 follows easily from the fact that the truth of a formula of
the form $\phi \Cond \psi$ or $\neg (\phi \Cond \psi)$ is independent of
the world, and depends only on the probability sequence.
}

Finally, I must show that (c) implies (b).  Suppose not.  Then there
exists a formula $\phi$ such that $\PS(\Lambda) \satp \phi$ but
$\PS(\Lambda) {\not{\sat}} \phi$.  Thus, there exists  $M \in \PS(\Lambda)$ and
valuation $V$ such that $(M,V) {\not{\sat}} \phi$.  
The proof in \cite{FHK} shows that if a formula is
satisfiable with respect to $\sat$ at all, then it is satisfiable in a
structure in $\PS(\Lambda)$ with a countable domain.  Thus, 
without loss of generality, $M$ has a countable domain.  But
then it immediately follows from Theorem~\ref{thm:equisatisfiable} that
$\PS(\Lambda) {\not{\satp}} \phi$.  \eprf

I next completely characterize reasoning in $\LCondfo^0$.  
I start by considering the fragment $\LCondfoo$ of $\LCondfo^0$
consisting of all formulas of the form $\phi \Cond \psi$
where $\phi$ and $\psi$ are closed first-order formulas.  Thus,
$\LCondfoo$ does not allow $\rarrow$ formulas to be universally
quantified.  
Consider the following rules:

\begin{description}
 \item[LLE.]
If $\vdashL \phi_1 \dimp \phi_2$, then from
 $\phi_1\Cond\psi$ infer
    $\phi_2\Cond\psi$ (left logical equivalence).
 \item[RW.]
If $\vdashL \psi_1 \rimp \psi_2$, then from
 $\phi\Cond\psi_1$ infer
    $\phi\Cond\psi_2$ (right weakening).
 \item[REF.] $\phi\Cond\phi$ (reflexivity).
 \item[AND.] From $\phi\Cond\psi_1$ and $\phi\Cond\psi_2$ infer
    $\phi\Cond \psi_1 \land \psi_2$.
 \item[OR.] From $\phi_1\Cond\psi$ and $\phi_2\Cond\psi$ infer
    $\phi_1\lor\phi_2\Cond \psi$.
 \item[CM.] From $\phi_1\Cond\phi_2$ and $\phi_1\Cond\psi$ infer
    $\phi\land \phi_2 \Cond \psi$ (cautious monotonicity).
\end{description}

This collection of rules has been called system $\sysPL$ \cite{KLM} or
\emph{the KLM properties}%
\footnote{$\Lambda$ is not usually mentioned explicitly, but it will be
useful to do so for the results of this paper.}
The rules are obvious analogues of axioms in $\AXCL$.
In particular, LLE is the analogue of R1, RW is the 
analogue of R2,  REF is the analogue C1, 
AND is the analogue of C2, OR is the analogue of C3, and CM is the analogue
of C4.  
Given a collection $\Delta$ of $\rarrow$ formulas, I write
$\sysPL \vdash \Delta \hra \phi\Cond\psi$ if $\phi\Cond\psi$ can be
derived from $\Delta$ using these rules.
A \emph{derivation from $\Delta$} consists of a sequence of steps of the
form  
$\Delta \hra \phi \Cond \psi$, where either (a) $\phi \Cond \psi \in
\Delta$, (b) $\phi = \psi$ (which can be viewed as an application of
the axiom REF), or (c) $\phi \Cond \psi$ follows from previous steps by
application of one of the rules in $\sysPL$.  All the rules above have
the form ``from $\phi_1 \rarrow \psi_1, \ldots \phi_n \rarrow \psi_i$'
infer $\phi \rarrow \psi$''; this can be viewed as an abbreviation for the
rule scheme ``from $\Delta \hra \phi_1 \Cond \psi_i, \ldots \Delta \hra
\phi_n \Cond \psi_n$ infer $\Delta \hra \phi \Cond \psi$'', with the
same $\Delta$ everywhere.  Although, for all these rules, the set
$\Delta$ is the same everywhere, later there will be 
rules where different sets $\Delta$ are involved.
I write $(M,V) \sat \Delta \hra \phi \Cond \psi$ if 
$(M,V) \sat \phi' \Cond \psi'$
for every formula $\phi' \Cond \psi' \in \Delta$ implies that $(M,V)
\sat \phi \Cond \psi$.  (For a formula $\phi \Cond \psi \in \LCondfoo$,
$\phi$ and $\psi$ are closed, so $(M,V) \sat \phi \Cond \psi$ iff $M \sat
\phi \Cond \psi$.   However, in $\LCondfo^0$ there are open
formulas, so the valuation $V$ plays a role.)  I write $\PS(\Lambda)
\sat \Delta \hra \phi \Cond \psi$ if $(M,V) \sat 
\Delta \hra \phi \Cond \psi$ for all PS structures $M$ and valuations $V$.
As usual, a rule is said to be \emph{sound} if it preserves truth (in this
case, with respect to all $(M,V)$); that is, if all the antecedents
hold with respect to $(M,V)$, then so does the conclusion.

\commentout{
\footnote{I remark that often these rules are generalized somewhat so
that an underlying propositional theory $\A$ is considered, 
and the rules LLE and RW consider provability with
respect to $\A$.  We can think of $\A$ as representing some domain
knowledge.  We could easily allow such domain knowledge here, but there
is really no loss of generality in not doing so.  I can capture any
formula $\phi$ that is intended to represent domain knowledge by 
the formula $\PBox \phi$.  (Recall that $\PBox \phi$ is an
abbreviation for $\neg \phi \Cond \false$.)  This guarantees that $\phi$
holds with probability 1, which is good enough for our purposes.}
}

\commentout{
I write $\Delta \sat \phi \Cond \psi$ if, for every 
propositional PS structure $M$ 
such that $M \sat \Delta$, $M \sat \phi \Cond \psi$ (where $M \sat
\Delta$ if $M \sat \phi' \Cond \psi'$ for each $\rarrow$ formula
$\phi' \Cond \psi' \in \Delta$).  We can similarly define 
$\Delta \satp \phi \Cond \psi$, by replacing $\sat$ with $\satp$.
(As we might expect, a propositional PS structure has the form
$(W,\pi,\P)$, where $W$ is a set of worlds, $\pi$ associates with each
world a truth assignment to primitive propositions, and $\P$ is a
probability sequence;
the semantics of formulas is obtained by the obvious modification of
the first-order semantics.

In the propositional case, it is well known that system {\bf P} is a
sound and complete axiomatization for $\PS$ under the $\sat$ semantics.
That is, if $\Delta$ is a
set of propositional $\rarrow$ statements, then 
$\Delta \vdash \phi \Cond \psi$ iff $\phi \Cond \psi$ is valid in PS
structures, using the $\sat$ semantics (see \cite{GMPfull,FrH5full}). 
Applying the arguments sketched in Theorem~\ref{thm:soundandcomplete1},
it is straightforward to show that system {\bf P} is also sound and complete  
for the $\satp$ semantics in the propositional case.
(Indeed, in \cite{FrH5full}, general techniques are given to show that
system {\bf P} is sound and complete for a wide range of semantics in
the propositional case.)  That is, we have the following result.
}

The following result is well known.

\begin{theorem}\label{thm:KLM} \cite{KLM,Geffner92} 
If $\Delta \union \{\phi \Cond \psi\} \subseteq \LCondfoo$, then
$\sysPL \vdash \Delta \hra \phi \Cond \psi$ iff
$\PS(\Lambda) \sat \Delta \hra \phi \Cond \psi$.
\end{theorem}

I want to extend this result from $\LCondfoo$ to $\LCondfo^0$, and to
the $\satp$ semantics as well as the $\sat$ semantics, so as to make it
applicable to reasoning about security protocols.  I actually extend it
to $\LCondfo^0 \union \Lfo$.
A collection $\Delta$ of formulas in $\LCondfo^0 \union \Lfo$ can be
written as 
$\Deltar \union \Deltafo$, where $\Deltar \subseteq \LCondfo^0$ and 
$\Deltafo \subseteq \Lfo$.  Consider the following 
strengthening of LLE:
\begin{description}
\item[LLE$^+$.] If $\vdash_{\Lambda \union \Deltafo} \phi \dimp \psi$,
then from $\Delta \hra \phi_1 \Cond \psi$ infer $\Delta \hra \phi_2
\Cond \psi$.
\end{description}
RW can be similarly strengthened to RW$^+$.  

Some rules from $\AXCL$ to deal with the universal
quantification are also needed,
specifically, variants of  
$\Lambda$-AX, F1, and F3, and another rule similar in 
spirit to F3:
\begin{description} 
\item[$\Lambda$-AX$^+$.] If $\vdash_{\Lambda \union \Deltafo} \phi$, 
then  $\Delta \hra \phi$.
\item[F1$^+$.] From $\forall x \phi$ infer $\phi[x/z]$,
where $z$ is a variable that does not appear in $\phi$. 
\item[F3$^+$.] If $x$ does not appear free in $\Delta$, then from
$\Delta \hra \phi$ infer $\Delta \hra \forall x \phi$.
\item[EQ.] If $x$ does not appear free in $\Delta$, $\phi$, or $\psi$,
and $\sigma$ is a first-order formula, then 
from $\Delta \union \{\sigma\} \hra \phi$ infer 
$\Delta \union \{\exists x \sigma\} \hra \phi$
(existential quantification).
\item[REN.] If $y_1, \ldots, y_n$ do not appear in $\phi$,
then from $\forall x_1, \ldots, x_n \phi$ infer
$\forall y_1, \ldots, y_n (\phi[x_1/y_1, \ldots, x_n/y_n])$
(renaming).
\end{description}

But these rules do not seem to suffice.  Intuitively, what is needed is 
a way to capture the fact that the domain is the same in all worlds 
In $\AXCL$, the one axiom that captures this is F5.  Unfortunately, F5 is
not expressible in $\LCondfo^0$.
To capture its effects in $\LCondfo^0$,
a somewhat more complicated rule seems necessary.
\begin{definition}
An \emph{interpretation-independent} formula $\phi$ is a first-order
formula that does not mention any constant, function, or predicate
symbols (and, thus, is a formula whose atomic predicates 
all are of the form $x=y$).
\end{definition}
The following rule can be viewed as a variant of the OR rule for
interpretation-independent formulas.
\begin{description}
\item[II.] If $\Delta \union \{\sigma_1\} \hra \phi$,
$\Delta \union \{\sigma_2\} \hra \phi$, and $\sigma_1$ and
$\sigma_2$ are interpretation-independent, then 
$\Delta \union \{\sigma_1 \lor \sigma_2\} \hra \phi$ 
(interpretation independence).
\end{description}

Let $\Pfop$ consist of $\sysPL$ (with LLE and RW replaced by LLE$^+$ and
RW$^+$, respectively) together with 
F1$^+$, F3$^+$, EQ, REN, and II.
\commentout{
I want to show that $\Pfop$ is sound and complete for derivations in
$\LCondfoo$.  To prove this, I need a preliminary lemma and definition.
$\Delta$ is \emph{closed under substitution} if $\phi \Cond \psi \in
\Delta$ implies that $(\phi \Cond \psi)[x/z] \in \Delta$, where $z$ is a
variable that does not appear in $\phi$ or $\psi$.  Let $\Pfom$ consist
of all the rules in $\Pfo$ other than SUB.  (Note that $\Pfom$ is just
system {\bf P}, except that I consider the first-order versions of LLE
and RW.)

\begin{lemma}\label{lem:sub} If $\Delta$ 
is closed under 
substitution, then $\Pfo \vdash \Delta \hra \phi \Cond \psi$ iff
$\Pfom \vdash \Delta \hra \phi \Cond \psi$.
\end{lemma}

\prf The ``only if'' direction is trivial.  For the ``if''
direction, suppose that $\Pfom \vdash \Delta \hra \phi \Cond \psi$.
I show by a straightforward induction on the length of the derivation that 
if $\Pfo \vdash \Delta \hra \phi' \Cond \psi'$ then 
$\Pfom \vdash \Delta \hra (\phi' \Cond \psi')[x/z]$.  I leave
details to the reader.  
\eprf
}

\begin{theorem}\label{thm:soundandcomplete}  
If $\Delta \union \{\phi\} \subseteq \LCondfo^0 \union \Lfo$, then
the following are equivalent: 
\begin{description}
\item[{\it (a)}] $\Pfop \vdash \Delta \hra \phi$;
\item[{\it (b)}] $\PS(\Lambda) \sat \Delta \hra \phi$;
\item[{\it (c)}] $\PS(\Lambda) \satp \Delta \hra \phi$.
\end{description}
\end{theorem}

\prf  The argument for soundness 
(that is, that (a) implies (c)) 
for the axioms and rules that also appear in $\sysPL$
is essentially done in the proof of Theorem~\ref{thm:soundandcomplete1};
the soundness of F1$^+$, F3$^+$, EQ, and REN is straightforward.
The soundness of II follows easily from the observation that, since there
is a fixed domain, if $\sigma_1$ and $\sigma_2$ are interpretation
independent and $(M,V) \sat \sigma_2 \lor \sigma_2$, then $(M,V) \sat
\sigma_1$ or $(M,V) \sat \sigma_2$. This would not be the case for a
formula such as $\bd_1 = \bd_2 \lor \bd_1=\bd_3$.  It could be that,
for every world $w$, $(M,V,w) \sat \bd_1 = \bd_2 \lor \bd_1 =
\bd_3$, with either $\bd_1= \bd_2$ being true in every world or
$\bd_1=\bd_3$ being true in every world.

The fact that (c) implies (b) follows
just as in the proof of Theorem~\ref{thm:soundandcomplete1},
using Theorem~\ref{thm:equisatisfiable}.  Thus, it remains to show that
(b) implies (a).   \shortv{See the full paper for
details. \eprf} 
\fullv{As usual, for completeness, it suffices to show that if
$\Pfop {\not{\vdash}} \Delta \hra \forall x_1 \ldots \forall x_n(\phi
\Cond \psi)$, then there is a 
structure $M \in \PS(\Lambda)$ and valuation $V$ such that $M \sat
\Delta$ and $(M,V) 
\sat \neg\forall x_1 \ldots \forall x_n ( \phi \Cond \psi)$.  
The idea is to reduce to
Theorem~\ref{thm:KLM} by transforming to a situation where the rules in
$\sysPL$ suffice.
Let $\DIST_k$ be the formula that says that there are at least $k$
distinct domain elements;
$$\exists x_1 \ldots \exists x_k (\land_{1 \le i < j \le k} x_i \ne x_j).$$
Note that $\DIST_k$ is interpretation-independent.
Let $\Delta_0 = \Delta$; let
$\Delta_{n+1} = \Delta_n \union \{ \DIST_{n+1}\}$ if
$\Pfo {\not{\vdash}} \Delta_{n} \union \{ \DIST_{n+1}\} \hra \phi
\Cond \psi$ and let  
$\Delta_{n+1} = \Delta_n \union \{\neg \DIST_{n+1}\}$ otherwise; 
finally, let $\Delta_\infty = \union_n \Delta_n$.  

An easy induction using II and $\Lambda$-AX$^+$ shows that $\Pfo
{\not{\vdash}} \Delta_n 
\hra \forall x_1 \ldots \forall x_n(\phi^* \Cond \psi^*)$ for all $n$, and
hence $\Pfo {\not{\vdash}} 
\Delta_\infty \hra \forall x_1 \ldots \forall x_n(\phi^* \Cond \psi^*)$.
Let $k^*$ be the largest $k$ such that $ \DIST_k \in \Delta_\infty$
(where take $k^* = \infty$ if, for all $k$,  $ \DIST_k \in
\Delta_\infty$).  Intuitively, $k^*$ will be the size of the domain in
the PS structure that we construct. 

By REN, I can assume without loss of generality that $x_1, \ldots, x_n$
do not appear in $\Delta_\infty$.  
Thus, from F3$^+$, it follows that 
$\Pfo {\not{\vdash}}  
\Delta_\infty \hra \phi \Cond \psi$.
A formula is a \emph{(complete) equality statement} for $x_1, \ldots, x_n$
if it is a conjunction of formulas of the form $x_i = x_j$ and $x_i \ne
x_j$, such that for all $1 \le 
i < j \le n$, either $x_i = x_j$ or $x_i \ne x_j$ is a conjunct.  Note
that a complete equality statement is interpretation-independent, and some
equality statement must be true of every valuation.  It thus follows
(using II and $\Lambda$-AX$^+$), that 
$\Pfo {\not{\vdash}} \Delta_\infty \union \{\sigma\} \hra \phi \Cond
\psi$, for some equality statement $\sigma$.  An equality statement $\sigma$
partitions $x_1, \ldots, x_n$ into equivalence classes, where it follows
from $\sigma$ that all the variables in each equivalence class are equal
to each other, but variables in two different equivalence classes are
not equal to each other.   Suppose that there are $h$ equivalence
classes. Clearly $h \le k^*$ (for otherwise $\sigma$ would be
inconsistent with $\dist_{k^*} \in \Delta_\infty$, so it would follow that
$\Pfo \vdash \Delta_\infty \union \{\sigma\} \hra \phi \Cond
\psi$).  Without loss of generality, I can assume that $x_1, \ldots,
x_h$ are in distinct equivalence classes (so that $\sigma$ implies that,
for all $j$, $x_1 = x_j \lor \ldots \lor x_h = x_j$).  Let $\phi'$ and
$\psi'$ be the result of replacing $x_j$ for $j > h$ by $x_i$ for $i \le
h$, where 
$\sigma \rimp x_i = x_j$.  It easily follows, using LLE$^+$ and RW$^+$,
that
$\Pfo {\not{\vdash}} \Delta_\infty \union \{\sigma\} \hra \phi' \Cond
\psi'$.  Let $y_i$, $1 \le i < k^*+1-h$ be fresh variables (where $k^*+1-h =
\infty$ if $k^* =\infty$) that do not
appear in $\Delta$, $\phi'$, or $\psi'$.  Let $\Delta_\infty^0 =
\Delta_\infty \union \{\sigma\}$; if $m > 0$, let $\Delta_\infty^{m} =
\Delta_\infty^{m-1} 
\union \{(\land_{1 \le j < m} y_m \ne y_j \land (\land_{1 \le j \le h}
y_m \ne x_j)\}$. Let $\Delta^\dag = \union_{i < k^*+1-h} \Delta_\infty^i$.  
I claim that 
$\Pfo {\not{\vdash}} \Delta^\dag \hra \phi' \Cond
\psi'$.  To show this, since proofs are finite, it suffices to show that
$\Pfo {\not{\vdash}} \Delta_\infty^m \hra \phi' \Cond \psi'$ for all $m
< k^*+1-h$.
I do this by induction on $m$.  For $m=0$, this is true by assumption.  
Suppose, by way of contradiction, that 
$\Pfo \vdash \Delta_\infty^m \hra \phi' \Cond \psi'$.
For $m > 0$, note that $\Delta_\infty^m$ has the form $\Delta_\infty
\union \{\sigma^0, \sigma^1, \ldots, \sigma^m\}$, where $y_m$ appears only in
$\sigma^m$.  It thus follows
from EQ that $\Pfo \vdash \Delta_\infty^m \union \{\sigma^0, \ldots,
\sigma^{m-1}, \exists y_m \sigma^m\} \hra
\phi' \Cond \psi'$. It is easy to see that $\DIST_m \rimp 
\exists y_m \sigma^m$ is valid.  Since 
$\DIST_m \in \Delta_\infty$, it follows that 
$\Pfo \vdash \Delta_\infty^{m-1} \hra
\phi' \Cond \psi'$, contradicting the inductive hypothesis.

Let $\bD = \{\bd_i: 1 \le i < k^*+1\}$ consist of
fresh constant symbols not in $\T \union \{\bc_1, \ldots, \bc_n\}$. 
Let an \emph{instantiation} of $\forall x_1 \ldots \forall x_n (\phi'
\Cond \psi')$ be a
formula in $\LCondfoo$ of the form $\phi'' \Cond \psi''$
that results by replacing each free variable in $\phi' \Cond
\psi'$ by some element of $\bD$.  For example, 
the instantiations of $\forall x \forall y (P(x,y) \Cond Q(y))$ are all
formulas
the form $P(\bd_i,\bd_j) \Cond Q(\bd_j)$.  
Let $\Delta^*$ be the result of replacing each formula $\forall x_1
\ldots \forall x_n( \phi' 
\Cond \psi')$ in $\Delta$ by all its instantiations, and replacing 
the free variables $x_1, \ldots, x_h, y_1, y_2, \ldots$ in the
formulas in $\Delta^\dag - \Delta$ by $\bd_1, \bd_2, \ldots$ respectively.
Let $\phi^*$ and $\psi^*$ be the result of replacing $x_1, \ldots, x_h$
in $\phi'$ and $\psi'$ by $\bd_1, \ldots, \bd_h$, respectively.
I claim that $\sysPLD {\not{\vdash}} \Delta^*_{\rarrow} \hra \phi^* \Cond
\psi^*$.
Suppose, by way of contradiction, that 
$\sysPLD \vdash \Delta^*_{\rarrow} \hra \phi^* \Cond \psi^*$.  
Then clearly $\Pfop \vdash \Delta^* \hra \phi^* \Cond \psi^*$.  
Let $\Delta_1$ be the result of replacing all occurrences of $\bd_1,
\bd_2, \ldots$  by $x_1, \ldots, x_h, y_1, y_2, \ldots$, respectively.
Then $\Pfop \vdash \Delta_1 \hra \phi' \Cond \psi'$ (simply replace
the constants $\bd_j$ by the appropriate variables in each line of the
derivation of $\phi^* \Cond \psi^*$).  Moreover,  $\Delta_1$ has the
form $\Delta_2 \union (\Delta^\dag - \Delta)$, and, by F1$^+$ and
$\Lambda$-AX$^+$, it easily follows that $\Delta \union (\Delta^\dag -
\Delta) \hra \phi' \Cond \psi$'; that is, $\Delta^\dag \hra \phi' \Cond
\psi'$.   This gives us the desired contradiction.

Since $\sysPLD {\not{\vdash}} \Delta^*_{\rarrow} \hra \phi^* \Cond
\psi^*$, it follows by Theorem~\ref{thm:KLM} that there exists a PS
structure $M \in \PS(\Lambda)$ and a valuation $V$ such that $M \sat
\Delta^*$ and $M \sat \phi^* \Cond \psi^*$.  Moreover, by standard
arguments (used, for example, in \cite{FHK}) the domain of
$M$ can be taken to be countable.  Since $ \DIST_k \in \Delta^*$ for all
$k < k^*$, 
and if $k^* < \infty$, $ \neg \DIST_{k^*} \in \Delta^*$, 
all the worlds in $M$ mus+t have domain size $k^*$ (where, if $k^* =
\infty$, the domain is countable).  Hence, I can assume without loss of
generality that all the worlds have the same domain, which can be take
to be $D = \{d_i: i < k^*\}$; moreover, 
$d_i$ can be taken to be the interpretation $\bd_i$ in each domain.  It
follows 
that $M \sat \Delta$, and $M {\not{\sat}} \phi \Cond \psi$, as desired.
\eprf
}

\subsection{Quantitative Reasoning}
The super-polynomial semantics just talks about asymptotic complexity.  
It says that for any polynomial $p$, the conclusion will hold
with probability greater than  $1  - 1/p(n)$ for sufficiently large $n$,
provided that the assumptions hold with sufficiently high probability,
where $n$ can be, for example, the security parameter.  While this
asymptotic complexity certainly gives insight into the security of a 
protocol, in practice, a system designer wants to achieve a certain
level of security, and needs to know, for example, how large to take the
keys in order  to achieve this.  In this section, I provide a more
quantitative semantics appropriate for such reasoning, 
and connect the qualitative and quantitative semantics.

The syntax of the quantitative language, which is denoted $\LCondfoq$, is
just like that of the 
qualitative language, except that, instead of formulas of the form $\phi
\Cond \psi$, there are  formulas of the form $\phi \Cond^r \psi$, where 
$r$ is a real number in $[0,1]$.  The
semantics of such a formula is straightforward:
$$\begin{array}{ll}
(M,V) \models \phi \Cond^r \psi \mbox{ if  there exists some $n^* \ge
0$ such that}\\ 
\mbox{for all $n
\ge n^*$, } {\Pr}_n(\intension{\psi}_{M,V}  \mid \intension{\phi}_{M,V}) \ge 1 -
r).
\end{array}$$
I define $\LCondfoq^0$ in the obvious way.  

For each of the axioms and rules in system $\sysPL$, there is a
corresponding sound axiom or rule in $\LCondfoq^0$:
\begin{description}
 \item[LLE$^q$.]
If $\vdash_{\Lambda \union \Deltafo} \phi_1 \dimp \phi_2$, then from
 $\Delta \hra \phi_1\Cond^r\psi$ infer
    $\Delta \hra \phi_2\Cond^r\psi$.
 \item[RW$^q$.]
If $\vdash_{\Lambda \union \Deltafo} \psi_1 \rimp \psi_2$,
then from 
 $\Delta \hra \phi\Cond^r \psi_1$ infer
    $\Delta \hra \phi\Cond^r\psi_2$. 
 \item[REF$^q$.] $\phi\Cond^0 \phi$ (reflexivity).
\item[AND$^q$.] From $\phi\Cond^{r_1}\psi_1$ and
$\phi\Cond^{r_2}\psi_2$ infer $\phi\Cond^{r_3} \psi_1 \land \psi_2$,
where $r_3 = \min(r_1 + r_2, 1)$.
\item[OR$^q$.] From $\phi_1\Cond^{r_1}\psi$ and
$\phi\Cond^{r_2}\psi_2$ infer $\phi_1 \lor \phi_2 \Cond^{r_3} \psi$, 
where $r_3 = \max(2r_1, 2r_2,1)$.
\item[CM$^q$.] From $\phi_1\Cond^{r_1}\phi_2$ and
$\phi_1\Cond^{r_2}\psi$ infer 
    $\phi\land \phi_2 \Cond^{r_3} \psi$, where $r_3 = \max(r_1+r_2,1)$.
\end{description}

Let $\Pfopq$ denote this set of rules, together with F1$^+$, F3$^+$, EQ,
REN, and II (all of which hold with no change in the quantitative
setting), and 
\begin{description}
\item[INC.] If $r_1 \le r_2$, then from $\phi \Cond^{r_1} \psi$
infer $\phi \Cond^{r_2} \psi$.
\end{description}

\begin{theorem} The rules in $\Pfopq$ are all sound.
\end{theorem}

\fullv{\prf
The soundness of the quantitative analogues of the rules in $\sysPL$ is
immediate from the proof of Theorem~\ref{thm:soundandcomplete1}.  
The soundness of remaining rules holds as it did before (since they are
unchanged).  
PS structure \eprf
}

I do not believe that $\Pfopq$ is complete, 
nor do I have a candidate complete axiomatization for the quantitative
language.   Nevertheless, as the proofs in \cite{DHMPR08} show, 
$\Pfopq$ suffices for proving many results of interest in security.
Moreover, as I now show, there is a deep relationship between $\Pfop$
and $\Pfopq$.  To make it precise, given a set of formulas $\Delta
\subseteq \LCondfo^0$, say that $\Delta' \subseteq \LCondfoq^0$ is a
\emph{quantitative instantiation} of $\Delta$ if, for every formula
$\phi \Cond \psi \in \Delta$, there is a bijection $f$ from $\Delta$ to
$\Delta'$ such that, for every formula $\phi \Cond \psi \in \Delta$,
there is a real number $r \in [0,1]$ such that $f(\phi \Cond \psi) =
\phi \Cond^r \psi$.  That is, $\Delta'$ is a quantitative instantiation
of $\Delta$ if each qualitative formula in $\Delta$ has a quantitative
analogue in $\Delta'$.

\fullv{
Although the proof of the following theorem is straightforward, it shows
the power of using of $\Pfop$.  Specifically, it shows} 
The following theorem shows that if $\phi \Cond \psi$ is derivable from 
$\Delta$ in $\Pfop$ then, for all $r \in [0,1]$, there exists
a quantitative instantiation $\Delta'$ of $\Delta$ such that $\phi
\Cond^r \psi$ is derivable from $\Delta'$ in $\Pfopq$.  Thus, if the
system designer wants security at level $r$ (that is, she wants to know
that the desired security property holds with probability at least
$1-r$), then if she has a qualitative proof of the result, she can
compute the strength with which her assumptions must hold in order for 
the desired conclusion to hold.  For example, she can compute how to set the
security parameters in order to get the desired level of security.  This
result can be viewed as justifying qualitative reasoning.  Roughly
speaking, it says that it is safe to avoid thinking about the
quantitative details, since they can always be derived later.
Note that this result
would not hold if the language allowed negation.  For example, even
if $\neg (\phi \Cond \psi)$ could be proved given some assumptions
(using the axiom system $\AXCL$), it would not necessarily follow that
$\neg (\phi \Cond^q \psi)$ holds, even if the probability of the
assumptions was taken arbitrarily close to one.  

\begin{theorem}\label{thm:quantitative} If $\Pfop \vdash \Delta \hra \phi \Cond
\psi$, then for all $r \in [0,1]$, there exists a quantitative
instantiation $\Delta'$ of $\Delta$ such 
that $\Pfopq \vdash \Delta' \hra \phi \Cond^q \psi$.  Moreover,
$\Delta'$ can be found in polynomial time, given the derivation of 
$\Delta \hra \phi \Cond \psi$.
\end{theorem}

\prf 
The existence of $\Delta'$ follows by a 
straightforward induction on the length of the derivation.  
\fullv{If it has
length 1, then either $\phi = \psi$, in which case it is an instance of
REF$^q$, or $\phi \Cond \psi \in \Delta$, in which case I simply take
$\Delta'$ such that it includes $\phi \Cond^r \psi$.  For the  inductive
step, the only nontrivial case is if $\phi \Cond \psi$ follows from
earlier steps by an instance of a rule of inference in $\Pfo$.  The
result then follows by a simple case analysis on the form of the rule.
For example, if the AND rule is used, then $\phi \Cond \psi$ has the
form $\phi \Cond \psi_1 \land \psi_2$, and there are shorter derivations
of $\Delta \hra \phi \Cond \psi_1$ and $\Delta \hra \phi \Cond \psi_2$.
\fullv{
Choose $s_1, s_2 \in [0,1]$ such that $s_1 + s_2 = r$%
\footnote{It suffices to take $s_1 = s_2 = r/2$, but there is an
advantage to having this greater flexibility; see the discussion after
the proof.}
}
By the induction hypothesis, there exist variants
$\Delta_1$ and $\Delta_2$ such that $\Pfoq \vdash \Delta_i \hra \phi
\fullv{
\Cond^{s_i} \psi_i$, for $i = 1, 2$.}
\shortv{\Cond^{r/2} \psi_i$, for $i = 1, 2$.}
  Let $\Delta_3$ be a quantitative
instantiation of $\Delta$ that dominates both
$\Delta_1$ and $\Delta_2$, in the sense that if $\phi' \Cond^{r_i} \psi'
\in \Delta_i$, for $i = 1,2,3$, then $r_i \ge \max(r_1,r_2)$.  
Then it is easy to 
\fullv{see that $\Pfoq \vdash \Delta_3 \hra \phi \Cond^{s_i} \psi_i$ for}
\shortv{see that $\Pfoq \vdash \Delta_3 \hra \phi \Cond^{r/2} \psi_i$ for}
$i = 1,2$. By 
AND$^q$, it easily follows that $\Pfoq \Delta_3 \hra \phi \Cond^{r}
\psi_1 \land \psi_2$.  The argument for all the other rules in $\Pfo$ is
similar. 
}
\fullv{This argument also shows that finding $\Delta'$ from the proof of
}
\shortv{The argument also shows that finding $\Delta'$ from the proof of }
$\Delta \hra \phi \Cond \psi$ just involves solving some simple linear
inequalities, which can be done in polynomial time.  
\shortv{See the full paper for details.}
\eprf

\fullv{
The proof of Theorem~\ref{thm:quantitative} gives even more useful
information to the system designer.  In general, there may be a number
of quantitative instantiations $\Delta'$ of $\Delta$ that give the
desired conclusion.  For example, as the proof shows, if the AND rule is
used in the qualitative proof, and we want the conclusion to hold at
level $r$, we must just choose $s_1$ and $s_2$ such that $\phi \Cond \psi_1$
and $\phi \Cond \psi_2$
hold at level $s_1$ and $s_2$, respectively.  If the system designer
finds it easier to satisfy the first formula than the
second (for example, the first may involve the length of the
key, while the second may involve the degree of trustworthiness of one
of the participants in the protocol), there may be an advantage in
choosing $s_1$ relatively small and $s_2$ larger.  As long as $s_1 + s_2 =
r$, the desired conclusion will hold. 
}

\paragraph{Acknowledgements:} I think Anupam Datta, John Mitchell,
Riccardo Pucella, and Arnab Roy for many useful discussions on 
applying conditional logic to security protocols.

\bibliographystyle{aaai}
\bibliography{z,joe,refs,riccardo2}
\end{document}